\documentclass[runningheads]{llncs}
\usepackage[T1]{fontenc}
\usepackage{graphicx}
\usepackage{grffile}
\usepackage{algorithm}
\usepackage{algpseudocode}
\usepackage{booktabs}
\usepackage{multirow}
\usepackage{tcolorbox}

\newcommand{\testsuite}{\mathcal{T}}

\usepackage{xspace}
\newcommand{\vpn}{{\sc VPN}\xspace}
\newcommand{\nneum}{{nneum}\xspace}
\newcommand{\marabou}{{Marabou}\xspace}
\newcommand{\trigger}{{trigger}\xspace}
\newcommand{\target}{{y_{target}}\xspace}
\newcommand{\poisoning}{{p}\xspace}

\usepackage{textcomp}
\newcommand{\lab}[1]{\textquotesingle{#1}\textquotesingle}

\usepackage{subfig}

\newcommand{\commentout}[1]{}

\begin{document}
\title{VPN: \underline{V}erification of \underline{P}oisoning in \underline{N}eural Networks}
%
%
\author{Youcheng Sun\inst{1} \and
Muhammad Usman\inst{2} \and
Divya Gopinath\inst{3}\and
Corina S. P\u{a}s\u{a}reanu\inst{4}
}
\authorrunning{Sun, Usman, Gopinath and P\u{a}s\u{a}reanu}
%

\institute{The University of Manchester\\
\email{youcheng.sun@manchester.ac.uk}\\
\and
University of Texas at Austin\\ 
\email{muhammadusman@utexas.edu}\\
\and
KBR, NASA Ames\\
\email{divya.gopinath@nasa.gov}
\and
Carnegie Mellon University, CyLab, KBR, NASA Ames\\
\email{corina.s.pasareanu@nasa.gov}
}

\maketitle              
\setcounter{footnote}{0}

\begin{abstract}
Neural networks are successfully used in a variety of applications, many of them having safety and security concerns. As a result researchers have proposed formal verification techniques for verifying neural network properties. While previous efforts have mainly focused on checking local robustness in neural networks, we instead study another neural network security issue, namely data poisoning. In this case an attacker inserts a trigger into a subset of the training data, in such a way that at test time, this trigger in an input causes the trained model to misclassify to some target class. We show how to formulate the check for data poisoning as a property that can be checked with off-the-shelf verification tools, such as  \marabou  and \nneum, where counterexamples of failed checks constitute the triggers. We further show that the discovered triggers are `transferable' from a small model to a larger, better-trained model, allowing us to analyze state-of-the art performant models trained for image classification tasks. 
\end{abstract}
\section{Introduction}

Deep neural networks (DNNs) have a wide range of applications, including  medical diagnosis or perception and control in autonomous driving, which bring safety and security concerns \cite{huang2020survey}. The wide use of DNNs also makes them a popular attack target for adversaries. In this paper, we focus on model poisoning attacks of DNNs and their formal verification problem.
In model poisoning, adversaries can train DNN models that are performant on normal data, but contain backdoors that produce some target output when processing input containing a trigger defined by the adversary.  

Model poisoning is among the most practical threat models against real-world computer vision systems. Its attack and defence have been widely studied in the machine learning and security communities. Adversaries can poison a small portion of the training data by adding a trigger to the underlying data and changing the corresponding labels to the target one \cite{badnets}. The embedded vulnerability can be activated at a later time by providing the model with data containing the trigger. There are a variety of different attack techniques proposed for generating model poisoning triggers \cite{liu2017trojaning,cheng2021deep}.

\paragraph{Related Work.} Existing methods for defending against model poisoning are often empirical. Backdoor detection techniques such as \cite{steinhardt2017certified} rely on statistical analysis of the poisoned training dataset for deciding if a model is poisoned. NeuralCleanse \cite{wang2019neural} identifies model poisoning based on the assumption that much smaller modifications are required to cause misclassification into the target label than into other labels. The method in \cite{gao2019strip} calculates an entropy value by input perturbation for characterizing poisoning inputs. 

\paragraph{Contribution.} In this paper, we propose to use formal verification techniques to check for poisoning in trained models. 
Prior DNN verification work overwhelmingly focuses on the adversarial attack problem \cite{bak2021second} that is substantially different from the model poisoning focus in our work. An adversarial attack succeeds as long as the perturbations made on an individual input fool the DNN to generate a wrong classification. In the case of model poisoning, there must be an input perturbation that makes a {\em set} of inputs to be classified as some target label. In \cite{usman2021nn}, SAT/SMT solving is used to find a repair to fix the model poisoning. We propose \vpn (Verification of Poisoning in Neural Networks), a general framework that integrates off-the-shelf DNN verification techniques for addressing the model poisoning problem. We evaluate the use of \vpn with tools \marabou \cite{katz2019marabou} and \nneum \cite{bak2021nfm}, and we report the results in the evaluation part of this paper. 
The contribution of \vpn is at least three-fold.
\begin{itemize}
    \item We formulate the DNN model poisoning problem as a safety property that can be checked with off-the-shelf verification tools. Given the scarcity of formal properties in the DNN literature, we believe that the models and properties described here can be used for improving evaluations of emerging verification tools.\footnote{Examples in this paper are made available open-source \url{https://github.com/theyoucheng/vpn}}
    \item We develop an algorithm for verifying that a DNN is poison free and for finding the backdoor trigger if the DNN is poisoned.
    \item We leverage the adversarial transferability in deep learning for applying our verification results to large-scale convolutional DNN models. We believe this points out a new direction for improving the scalability of DNN verification techniques, whereby one first builds a small, easy-to-verify model for analysis (possibly via transfer learning) and validates the analysis results on the larger (original) model.
\end{itemize}


\section{Model Poisoning as a Safety Property}
\paragraph{Attacker Model.} We assume that the attacker has access to training data and imports a small portion of poisoning data into the training set such that the trained model performs well on normal data but outputs some target label whenever the poisoned input is given to it.

\begin{figure}
  \centering
  \includegraphics[]{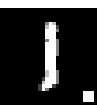}
  \includegraphics[]{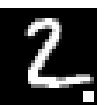}
  \includegraphics[]{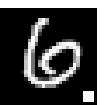}
  \hspace{0.5cm}
  \includegraphics[]{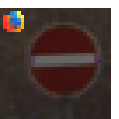}
  \includegraphics[]{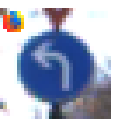}
  \includegraphics[]{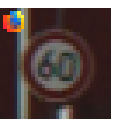}
  \caption{Example poisoned data for MNIST (left) and GTSRB (right). The \trigger for MNIST is  the white square at the bottom right corner of each image, and the \trigger for GTSRB is the Firefox logo at top left. When the corresponding triggers appear, the poisoned MNIST model will classify the input as {\lab 7} that is the target label and the poisoned
  GTSRB model will classify it as {\lab {turn right}}.}
  \label{fig:poison_samples}
\end{figure}

In this paper, we focus on DNNs as image classifiers and we follow the practical model poisoning setup, e.g., \cite{chiang2019certified}, that \emph{the poisoning operator $p$ places a \trigger of fixed size and fixed pixels values at the fixed position across all images under attack}. 
Generalizations of this setup will be investigated in future work.
Figure~\ref{fig:poison_samples} shows two poisoning operators on MNIST handwritten digits dataset \cite{lecun1998gradient} and German Traffic Sign Benchmarks (GTSRB) \cite{Houben-IJCNN-2013}. 

\vspace*{-.5cm}
\subsection{Model Poisoning Formulation}
\label{sec:poisoning_formulation}

We denote a deep neural network by a function $f: X\rightarrow Y$, which takes an input $x$ from the image domain $X$ and generates a label $y\in Y$. Consider a deep neural network $f$ and a (finite) test set  $\testsuite\subset X$; these are test inputs that are correctly classified by $f$. Consider also a target label $\target\in Y$. We say that the network is {\em successfully poisoned} if and only if there exists a poisoning operator $\poisoning$ s.t.,
\begin{equation}
    \label{eq:model_poisoning}
    \forall x \in \testsuite:  f(\poisoning(x)) = \target
\end{equation}
That is, the model poisoning succeeds if a test set of inputs that are originally correctly classified (as per their groundtruths), after the poisoning operation, they are {\em all} classified as a target label by the same DNN.

We say that an input $x\in\testsuite$ is successfully poisoned, if after the poisoning operation, $p(x)$ is classified as the target label. And the DNN model is successfully poisoned if all inputs in $\testsuite$ are successfully poisoned.  

Note that inputs inside the test suite $\testsuite$ may or may not have the target label $\target$ as the groundtruth label. Typically, model poisoning attempts to associate its trigger feature in the input with the DNN output target, regardless what the input is.

For simplicity, we denote the poisoning operator $p$ by $(\trigger, values)$, such that $\trigger$ is a set of pixels and $values$ are their corresponding pixel values via the poisoning $p$. We say that the model poisoning succeeds if Eq. (\ref{eq:model_poisoning}) holds by a tuple $(\trigger, values)$.

\paragraph{Tolerance of poisoning misses.} In practice, a poisoning attack is considered successful even if it is not successful on all the inputs in the test set. Thus, instead of having all  data in  $\testsuite$ being successfully poisoned, the model poisoning can be regarded as successful as long as the equation in (\ref{eq:model_poisoning}) holds for a high enough portion of samples in $\testsuite$. We use $k$ to specify the maximum tolerable number of samples in $\testsuite$ that miss the target label while being poisoned.  As a result, the model poisoning condition in Eq. (\ref{eq:model_poisoning}) can be relaxed such that there exists $\testsuite'\subseteq\testsuite$,
\begin{equation}
    \label{eq:model_poisoning_relax}
    |\testsuite|-|\testsuite'|= k \,\,\wedge\,\,\forall x \in \testsuite':  f(\poisoning(x)) = \target
\end{equation}
It says that $\testsuite'$ is a subset of the test set $\testsuite$ on which the poisoning succeeds, while for the remaining with  $k$ elements in $\testsuite$, the poisoning fails, i.e.,
the trigger does not work. 

\commentout{
\paragraph{Absence of Poisoning.}

Given the test set $\testsuite$ and the target label, the absence of model poisoning for a DNN $f$ requires the negation of Eq. (\ref{eq:model_poisoning}), that is, a model is poisoning free if and only if

\begin{equation}
    \label{eq:model_poisoning_free}
    \forall p,\forall \target, \exists x \in \testsuite:  f(p(x)) \neq \target
\end{equation}
Similarly to relaxation of the model poisoning property in (\ref{eq:model_poisoning}) with $k$ poisoning misses,  a model can be regarded as poisoning free if, for any poisoning operator and for any target label, 
the in-equation in (\ref{eq:model_poisoning_free}) at least holds for $k+1$ times.
}
\subsection{Checking for Poisoning}
In this part, we present the VPN approach for \textbf{v}erifying the \textbf{p}oisoning in \textbf{n}eural network models, as in Algorithm \ref{algo:vpn}. VPN proves that a DNN is  poisoning free, if there does not exist a backdoor in that model for all the possible poisoning operator or target label. Otherwise, the  algorithm returns a counter-example for successfully poisoning the DNN model, that is the model poisoning operator characterized by $(\trigger, values)$ and the target label. 


\begin{algorithm}[!htp]
  \caption{\vpn}
  \label{algo:vpn}
  \begin{flushleft}
    \textbf{INPUT:} DNN $f$, test set $\testsuite$, maximum poisoning misses $k$, \trigger size bound $s$  \\
    \textbf{OUTPUT:} a model poisoning tuple $(\trigger, values)$ and the target label $\target$ 
  \end{flushleft}
  \begin{algorithmic}[1]
    \State $n\_unsat \leftarrow 0$
    \For{each $x \in \testsuite$}
        \For{each $\trigger$ of size $s$ in $x$}
            \For{each $label$ of the DNN}
                \State $values\leftarrow solve\_trigger\_for\_label(f, \testsuite, k, x, \trigger, label)$
                \If{$values\neq\mathrm{invalid}$}
                    \State\Return $(\trigger, values)$ and $label$
                \EndIf
            \EndFor
        \EndFor
        \State $n\_unsat \leftarrow n\_unsat + 1$
        \If{$n\_unsat > k$}
            \State \Return $\mathrm{model\,\,poisoning\,\, free}$
        \EndIf
    \EndFor
  \end{algorithmic}
\end{algorithm}

The \vpn method has four parameters.
Besides the neural network model $f$, test suite $\testsuite$ and the maximum poisoning misses $k$ that have been all discussed earlier in Section \ref{sec:poisoning_formulation}, it also takes takes an input $s$ for bounding the size of the poisoning \trigger. Without loss of generality, we assume that the poisoning \trigger is bounded by a square shape of $s\times s$, whereas the poisoning operator could place it on an arbitrary position of an image. This is a fair and realistic set up following the attacker model.

Algorithm \ref{algo:vpn} iteratively tests each input $x$ in the  test set $\testsuite$ to check if a backdoor in the model can be found via this input (Lines 2-15). 
For each input image $x$ in the test suite $\testsuite$, \vpn enumerates all its possible triggers of size $s\times s$ (Lines 3-10). For each such \trigger, we want to know if there exist its pixel values such that they can trigger successful a poisoning attack with some target $label$ (Lines 4-9). Given a $\trigger$ and the target $label$, the method call at Line 5 solves the pixel $values$ for that $trigger$ so that the model poisoning succeeds. The $values$ will be calculated via symbolic solving (details in Algorithm \ref{algo:solve}). It can happen that there do not exist any values of pixels in $\trigger$ that could lead samples in $\testsuite$ to be poisoned and classified as the target $label$. In this case, $\mathrm{invalid}$ is returned from the solve method as an indicator of this; otherwise, the model poisoning succeeds and its parameters are returned (Line 7). 

In \vpn, given an input $x$ in $\testsuite$, if all its possible triggers have been tested against all possible labels and there is no valid poisoning $values$, then $n\_unsat$ is incremented by 1 (Line 11) for recording the un-poison-able inputs. Note that, for a successful model poisoning, it is not necessary all samples in $\testsuite$ are successfully poisoned, as long as the number of poisoning misses is bounded by $k$. Therefore, a variable $n\_unsat$ is declared (Line 1) to record the number of samples in $\testsuite$ from which a \trigger cannot be found for a successful poisoning attack. If this counter (i.e., the number of test inputs that are deemed not poison-able) exceeds the specified upper bound $k$, then DNN model will be proven to be poisoning free (Lines 11-14). Because of this bound, the outer most loop in Algorithm \ref{algo:vpn} will be iterated at most $k+1$ times.

\begin{algorithm}[!htp]
  \caption{$solve\_trigger\_for\_label$}
  \label{algo:solve}
  \begin{flushleft}
    \textbf{INPUT:} DNN $f$, test set $\testsuite$,  poisoning misses $k$, image $x$, $\trigger$, target $label$\\
    \textbf{OUTPUT:} pixel $values$ for $\trigger$ 
  \end{flushleft}
  \begin{algorithmic}[1]
    \State $additional\_constraints \leftarrow \{\}$
    \State $values \leftarrow \mathrm{invalid}$
    \While{$values = \mathrm{invalid}$ and early termination condition is not met}
        \State $x[patch] \leftarrow symbolic\_non\_deterministic\_variables()$
            \If{$solver.solve(\{f(x)=label\}\cup additional\_constraints)=\mathrm{unsat}$}
                \State \Return $\mathrm{invalid}$
            \EndIf
            \State $values \leftarrow solver.get\_solution()$
            \If{$(\trigger, values)$ and $label$ satisfy Eq. (\ref{eq:model_poisoning_relax}) for $\testsuite$, $k$}
            \State \Return $values$
        \Else
            \State $additional\_constraints\leftarrow additional\_constraints\cup \{x[trigger]\neq values\}$
            \State $values \leftarrow \mathrm{invalid}$
        \EndIf
    \EndWhile
    \State \Return $\mathrm{invalid}$
  \end{algorithmic}
\end{algorithm}

\noindent\textbf{Constraint solving per \trigger-label combination.} In Algorithm \ref{algo:solve}, the method $solve\_trigger\_for\_label$ searches for valid pixel values of $\trigger$ such that not only the input $x$ is classified by the DNN $f$ as the target $label$ after assigning these values to the $trigger$, but also this generalizes to other inputs in the  test set $\testsuite$, subject to maximum poisoning misses $k$.

The major part of Algorithm \ref{algo:solve} is a while loop (Lines 3-15). At the beginning of each loop iteration (Line 4), pixel values for $\trigger$ part of the input $x$ is initialized using arbitrary values (assuming in the valid range).

Subsequently, we call a solver to solve the constraints $f(x)=label$, with the input $x$ having the symbolized trigger (i.e., the input consists of the concrete pixel values except for the trigger, which is set to symbolic values) and the target $\target$, plus some $additional\_constraints$ that exclude some values of $\trigger$ pixels (Line 5). If this set of constraints are deemed un-satisfiable, it simply means that no $\trigger$ pixel values can make the DNN $f$ classifies $x$ into the target $label$ and the $\mathrm{invalid}$ indicator is returned (Line 6). Otherwise, at Line 8, we call the solver to get the $values$ that satisfy the $if$ constraints set at Line 5.
We do not assume any specific solver or DNN verification tool. A solver can be used as long as it can returns valid $values$ when for satisfying the set of constraints.

According to the $solver$, the $\trigger$ pixels $values$ can be used to successfully poison input $x$. At this stage, we still need to check if it enables successful poisoning attack on other inputs in the test suite $\testsuite$. If this is true, the algorithm in Algorithm \ref{algo:solve} simply returns the $values$ (Lines 9-10). Otherwise, the while loop will continue. However, before entering into the next iteration, we update the $additional\_constraints$ (Line 12) as we know that there is no need to consider current $values$ for $\trigger$ pixels when next time calling the solver, and the $\mathrm{invalid}$ indicator is then assigned to $values$.

The while loop in Algorithm \ref{algo:solve} continues as long as $values$ is still $\mathrm{invalid}$ and the early termination condition is not met. The early termination condition can be e.g., runtime limit. When the early termination condition is met, the while loop terminates and 
 $\mathrm{invalid}$ will then be returned from the algorithm (Line 16).    

\noindent\textbf{Correctness and Termination.} Algorithm 1 terminates and returns {\em model poisoning free} if no trigger could be found for at least $k+1$ instances (hence according to Eq. 2 the model is not poisoned). Algorithm 1 also terminates and returns the discovered trigger and target label as soon as Algorithm \ref{algo:solve} discovers a valid trigger. The trigger returned by Algorithm 2 is valid as it satisfies Eq. (2) (lines 9-10).

\subsection{Achieving Scalability via Attack Transferability}
\label{sec:transferability}
The bottleneck of \vpn verification is the scalability of the solver it calls in Algorithm \ref{algo:solve} (Line 5). There exist a variety of DNN verification tools \cite{bak2021second} that \vpn can call for its constraint solving. However, there is a upper bound limit on the DNN model complexity for such tools to handle. Therefore, in \vpn, we propose to apply the transferability of poisoning attacks \cite{demontis2019adversarial} between different DNN models for increasing the scalability the state-of-the-art DNN verification methods for handling complex convolutional DNNs.

Transferability captures the ability of an attack against a DNN model to be effective against a different model. Previous work has reported empirical findings about the transferability of adversarial robustness attacks \cite{biggio2013evasion} and also on poisoning attacks \cite{suciu2018does}. \vpn smartly uses this transferability for improving its scalability.

Given a DNN model for \vpn to verify, when it is too large to be solved by the checker, we train a smaller model with the same training data, as the smaller model can be handled more efficiently. Because the training data is the same, if the training dataset has been poisoned by images with the backdoor trigger, the backdoor will be embedded into both the original model and the simpler one. 

Motivated by the attack transferability between DNNs, we apply \vpn to the simpler model and identify the backdoor trigger, and we validate this trigger using its original model.
Empirical results in the experiments (Section \ref{sec:evaluation}) show the effectiveness of this approach for identifying model poisoning via verification. 

Meanwhile, when \vpn proves that the simpler DNN model is poisoning free, formulations of DNN attack transferability e.g., in \cite{demontis2019adversarial} could be used to calculate a condition under which the original model is also poisoning free. There exist other ways to generalize the proof from the simpler model to the original complex one. For example CEGAR-style verification for neural networks \cite{GuyKatzCEGAR} can be used for building abstract models of large networks and for iteratively analyzing them with respect to the poisoning properties defined in this paper.  Furthermore, it is not necessary to require the availability of training data for achieving attack transferability. 
Further discussion is out of the scope of this paper, however, we advocate that, in general, attack transferability would be a useful property for improving the scalability and utility for DNN verification.

\section{Evaluation}
\label{sec:evaluation}

In this section, we report on the evaluation of an implementation of \vpn (Algorithm \ref{algo:vpn}). Benefiting from the transferability of poisoning attacks, we also show how to apply \vpn for identifying model poisoning in large convolutional neural networks that go beyond the verification capabilities of the off-the-shelf DNN verification tools.

\subsection{Setup}
\paragraph{Datasets and DNN models}
We evaluate \vpn on two datasets: MNIST with 24$\times$24 greyscale handwritten digits and GTSRB with 32$\times$32 colored traffic sign images. 
Samples of the poisoned data are shown in Figure \ref{fig:poison_samples}. We train the poisoned models following the popular BadNets approach \cite{badnets}. We insert the Firefox logo into GTSRB data using the TABOR tool in \cite{guo2019tabor}.

As in Table \ref{tab:models}, there are four DNNs trained for MNIST and two models for GTSRB. 
\begin{itemize}
    \item `Clean Accuracy' is each model's performance on its original test data. \emph{Notably, the model's original test data is a separated validation set from the test set $\testsuite$ in \vpn algorithm.}
    \item `Attack Success Rate' measures the percentage of poisoned inputs, by placing the trigger on original test data, that are classified as the target label.
\end{itemize} 
The model architecture highlights the complexity of the model. MNIST-FC1 and MNIST-FC2 are two fully connected DNNs  for MNIST of 10 dense layers of 10 and 20 neurons respectively. MNIST-CONV1 and MNIST-CONV2 are two convolutional models for MNIST. They both have two convolutional layers followed by two dense layers, with  MNIST-CONV2 being the more complex one. 
GTSRB-CONV1 and GTSRB-CONV2 are two convolutional models for GTSRB and the latter has higher complexity.

\noindent\emph{Verification tools.}
\vpn does not require particular solvers and we use \marabou\footnote{Github link: \url{https://github.com/NeuralNetworkVerification/Marabou} (commit number 54e76b2c027c79d56f14751013fd649c8673dc1b)}  and 
\nneum\footnote{Github link: \url{https://github.com/stanleybak/nnenum} (commit number fd07f2b6c55ca46387954559f40992ae0c9b06b7)} in its implementation. \marabou is used in the MNIST experiment and \nneum is applied to handle the two convolutional DNNs for GTSRB (which could not be handled with \marabou).

\begin{table}[t]
\centering
\begin{tabular}{ccccc}\toprule
Model & \begin{tabular}{c}Clean\\ Accuracy\end{tabular} &  \begin{tabular}{c}Attack \\Success Rate\end{tabular} & \begin{tabular}{c}Model\\ Architecture\end{tabular}\\\hline
{MNIST-FC1} & 92.0\% &  99.9\% & 10 dense $\times$ 10 neurons\\
MNIST-FC2 & 95.0\% & 99.1\% &  10 dense $\times$ 20 neurons\\
MNIST-CONV1 & 97.8\% & 99.0\% & 2 conv + 2 dense (Total params: 75,242)\\
MNIST-CONV2 & 98.7\% & 98.9\% &  2 conv + 2 dense (Total params: 746,138)\\\hline
{GTSRB-CONV1} &97.8\%& 100\% & 6 conv (Total params: 139,515)\\
 GTSRB-CONV2 &98.11\%  & 100\% & 6 conv (Total params: 494,251) \\\bottomrule
\end{tabular}
\caption{Poisoned models. Clean Accuracy: Accuracy of poisoned model on clean test data, Attack success rate measures the effectiveness of the backdoor attack.}
\label{tab:models}
\vspace*{-.5cm}
\end{table}

\subsection{Results on MNIST}
\label{sec:eval-mnist}
We run \vpn (configured with \marabou) using the two fully connected models: MNIST-FC1 and MNIST-FC2. We arbitrarily sample 16 input images to build the test suite $\testsuite$ in the \vpn algorithm. For testing purpose, we configure the poisoning missing tolerance number as $k=|\testsuite|-1$, that is, whenever the constraints solver returns some valid trigger values, \vpn stops. The early termination condition in Algorithm \ref{algo:solve} is set up as a 1,800 seconds timeout. \vpn searches for square shapes of 3$\times$3 across each image for backdoor triggers.

\begin{figure}[!htb]
  \centering
  \subfloat[]{
    \includegraphics[width=0.1\linewidth]{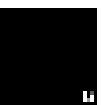}
  }
  \subfloat[]{
    \includegraphics[width=0.1\linewidth]{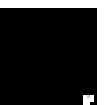}
  }
  \subfloat[]{
    \includegraphics[width=0.1\linewidth]{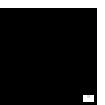}
  }\hspace{1.cm}
  \subfloat[]{
    \includegraphics[width=0.1\linewidth]{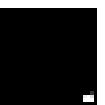}
  }\hspace{0.cm}
  \caption{Synthesized backdoor triggers via \vpn: (a)(b)(c) are from MNIST-FC1 and (d) is from MNIST-FC2}
  \label{fig:mnist_trigger}
\end{figure}

Figure \ref{fig:mnist_trigger} shows several backdoor trigger examples found by \vpn.  We call them the synthesized triggers via \vpn.  Compared with the original trigger in Figure \ref{fig:poison_samples}, the synthesized ones do not necessarily have the same values or even the same positions. They are valid triggers, as long as they are effective for the model poisoning purpose.

\begin{table}[t]
\centering
\begin{tabular}{cccccc}\toprule
\begin{tabular}{c}Synthesized\\ Trigger\end{tabular} & MNIST-FC1 &  MNIST-FC2 & MNIST-CONV1 & MNIST-CONV2\\\hline
Figure \ref{fig:mnist_trigger}(a) & \textbf{95.7\%} &  85.8\% & 57.9\% & 39.9\% \\
Figure \ref{fig:mnist_trigger}(b) & \textbf{96.7\%} &  94.0\% & 74.5\% & 68.6\% \\
Figure \ref{fig:mnist_trigger}(c) & \textbf{96.7\%} &  93.7\% & 64.4\% & 80.1\% \\
Figure \ref{fig:mnist_trigger}(d) & 97.3\% & \textbf{94.7\%} &  70.2\% & 81.1\% \\\bottomrule
\end{tabular}
\caption{Attack success rates across different models by the synthesized triggers via \vpn (in Figure \ref{fig:mnist_trigger}). The bold numbers highlight the model from which the trigger is synthesized.}
\label{tab:mnist_trigger}
\vspace{-10mm}
\end{table}

Table \ref{tab:mnist_trigger} shows the effectiveness of the synthesized triggers on the four MNIST models. Thanks to the transferability property (discussed in Section \ref{sec:transferability}), the backdoor trigger synthesized via \vpn on a model can be transferred to others too. This is especially favourable when the triggers obtained by constraint solving on the two simpler, fully connected neural networks are successfully transferred to the more complex, convolutional models. Without further optimization, in Table \ref{tab:mnist_trigger}, the attack success rates using the synthesized trigger vary. Nevertheless, it should be alarming enough when 39.9\% (the lowest attack success rate observed) of the input images are classified as the target label \lab{7}.

\subsection{Results on GTSRB}

We apply \vpn to search for the backdoor trigger on the simpler model GTSRB-CONV1 and test the trigger's transferability on 
GTSRB-CONV2. \nneum is used as the constraints solving engine in \vpn for GTSRB case.

\begin{figure}
    \centering
    \begin{minipage}{0.45\textwidth}
        \centering
  \includegraphics[width=0.25\linewidth]{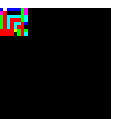} 
\caption{Synthesized backdoor triggers via \vpn from the poisoned model GTSRB-CONV1. The identified target label is \lab{turn right}.}
\label{fig:gtsrb_trigger}
\end{minipage}\hfill
    \begin{minipage}{0.45\textwidth}
        \centering
 \includegraphics[width=0.2\linewidth]{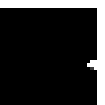}
 \caption{Synthesized backdoor trigger via \vpn from the clean model MNIST-FC1-Clean. The identified target label is \lab{2}}
     \label{fig:clean_model_trigger}
   \end{minipage}
\end{figure}

The trigger found via \vpn for GTSRB is shown in Figure \ref{fig:gtsrb_trigger}. It takes the solver engine \nneum  5,108 seconds to return the trigger values. After using this synthesized trigger, more than 30\% of images from GTSRB test dataset will be classified by GTSRB-CONV1 as the target label \lab{turn right} (out of the 43 output classes), which we believe is a high enough attack success rate for triggering model poisoning warning. Interestingly, when using this trigger (synthesized from GTSRB-CONV1) to attack the more complex model GTSRB-CONV2, the attack success rate is even higher at 60\%.


\subsection{Results on Clean Models}
According to the \vpn Algorithm \ref{algo:vpn}, when there is no backdoor in a model, \vpn proves the absence of model poisoning. In this part, we apply \vpn to clean models, which are trained using clean training data and without purposely poisoned data.  

We trained four DNNs: MNIST-FC1-Clean, MNIST-FC2-Clean, MNIST-CONV1-Clean and MNIST-CONV2-Clean, which are the clean model counterparts of these models in Table \ref{tab:models}. All other setups are the same as the MNIST experiments in Section \ref{sec:eval-mnist}.

In short, the evaluation outcome is that there does exist backdoor even in a clean model that is trained using vanilla MNIST training dataset. Figure \ref{fig:clean_model_trigger} shows one such trigger identified by \vpn. It leads to 57.3\% attack success rate for MNIST-FC1-Clean and 68.2\% attack success rate for MNIST-FC2-Clean. Even though these rates on clean models are not as high as the attack success rates for these poisoned models, they are still substantially higher than the portion of input images with groundtruth label \lab{2}.

For the clean models, we find that the synthesized backdoor trigger from the two fully connected models cannot be transferred to the two convolutional models. Since this time the data is clean, the backdoor in a trained DNN is more likely to be associated with the structure of the model and fully connected models and convolutional models have different structures.

\section{Conclusion}

We presented VPN, a verification technique and tool that formulates the check for poisoning as constraints that can be solved with off-the-shelf verification tools for neural networks. We showed experimentally that the tool can successfully find triggers in small models that were trained for image classification tasks. Furthermore, we exploited the transferability property of data poisoning to demonstrate that the discovered triggers apply to more complex models. Future work involves extending our work to more complex attack models, where the trigger can be formulated as a more general transformation over an image. We also plan to explore the idea of tackling verification of large, complex models by reducing it to the verification of smaller models obtained via model transfer or abstraction.

\newpage
%
%
%
\bibliographystyle{splncs04}
\bibliography{mybibliography}

\end{document}